\documentclass{edp-jp4}
\usepackage{graphicx}
\begin{document}

\title{The GAIA astrometric survey of extra-solar planets} 
\author{M. G. Lattanzi}\address{Osservatorio Astronomico di Torino, 
10025 Pino Torinese, Italy}
\author{S. Casertano}\address{Space Telescope Science Institute, 
Baltimore, MD 21218, USA}
\author{A. Sozzetti}\address{University of Pittsburgh, Dept. 
of Physics \& Astronomy, Pittsburgh, PA 15260, USA}\sameaddress{,1}
\author{A. Spagna}\sameaddress{1}
\maketitle
\begin{abstract} 
The ESA Cornerstone Mission GAIA, to be launched prior to 2012 and with 
a nominal lifetime of 5 years, will improve the accuracy of Hipparcos 
astrometry by more than two orders of magnitude. 

GAIA high-precision global astrometric measurements will provide deep insights 
on the science of extra-solar planets. The GAIA contribution is primarily 
understood in terms of the number and spectral type of targets available 
for investigation, and characteristics of the planets to be searched for.
Several hundreds of thousands of solar-type stars (F-G-K) within a sphere 
of $\sim 200$ pc centered on our Sun will be observed. GAIA will be 
particularly sensitive to giant planets (M$_\mathrm{p}$$\sim$M$_\mathrm{J}$) 
on wide orbits, up to periods twice as large as the mission duration, the 
potential signposts of the existence of rocky planets in the Habitable Zone. 
Thousands of new planets might be discovered, and a significant fraction 
of those which will be detected will have orbital parameters measured to 
better than $30\%$ accuracy.  By measuring to a few degrees 
the relative inclinations of planets in multiple systems with favorable 
configurations, GAIA will also make measurements of unique value towards a 
better understanding of the formation and evolution processes of planetary 
systems. 
\end{abstract}
%
\section{Introduction}

The present catalogue of candidate extra-solar planets discovered by 
radial velocity surveys (see 
for example~\cite{butler00}) totals today 66 objects having minimum mass 
M$\sin i \le 13$ M$_\mathrm{J}$ (where M$_\mathrm{J}$ is the mass of Jupiter), 
the so-called deuterium-burning threshold. 
Orbital periods span a range between a few days and 
$\sim 7$ years, but $\sim 80\%$ of these objects revolves around 
the parent star on orbits with semi-major axis $a < 1$ AU, 
well outside the ice condensation zone. 
Orbital eccentricities are usually higher than in our Solar 
System, up to $\sim 0.9$ for HD 80606~\cite{naef01}. Seven stars harbor
multiple systems, formed by either two or three planets~\cite{butler99,
mayor00,marcy01a,fischer02}, or a planetary mass object and a probable brown 
dwarf~\cite{udry00,marcy01b,els01}.

Except for the case of HD 209458~\cite{henry00}, 
all low-mass companions to solar type stars 
having $M_\mathrm{p} \leq 13 M_\mathrm{J}$ have been 
classified as extra-solar planets solely on the 
basis of their small projected masses, and thus, under the reasonable assumption 
of randomly oriented orbital planes on the sky, small true masses. But, 
some of them 
may not be planets at all, as $\sim 1/5$ of them have M$\sin i > 5$ 
M$_\mathrm{J}$. As a matter 
of fact, today the true nature of these objects is still matter of ongoing debates 
among the scientific community: for example, planet (and brown dwarf) candidates, 
and stellar binaries have remarkably similar 
orbital elements distributions~\cite{stepinski01}, but the mass 
functions in the two cases are strikingly different~\cite{mayor01}.

Clearly, our present
understanding of the origin of planetary systems is still limited, and 
more measurements will be needed in order
to be able to discriminate among models. To go beyond a simple 
{\bf\underline{Catalogue}} of extra-solar planets,  
{\bf\underline{Classification}} will have to be made on the basis 
of the knowledge of their true masses, shape and alignment of the orbits, 
structure and composition of the atmospheres. 
The dependence of planetary frequencies with age 
and metallicity will have to be understood. 
Finally, important issues on planetary 
systems evolution, such as coplanarity and long-term stability, will have to 
be addressed. But, the big picture will not be complete without crucial 
{\bf\underline{New Discoveries}}. The existence of giant planets orbiting 
on Jupiter-LIKE orbits (4-5 AU, or more) will have to be established. Such 
objects are the signposts for the discovery of rocky planets orbiting in 
regions closer to their parent stars, maybe even inside 
the star's Habitable Zone (for its definition, see Section~\ref{zone}). The 
proof of the existence of Earth-LIKE planets would be an extraordinary step 
towards the ultimate goal of the discovery of extra-terrestrial life.

\section{The role of GAIA}

GAIA's ability in detecting and measuring planets is twofold, 
it will impact both future planet discoveries as well as provide information 
of great value for a compute classification of planetary systems and overall 
assessment of the correct theories of planet formation and evolution. 
In particular, the uniqueness of the GAIA contribution 
to the science of extra-solar planets is better understood in terms of $1)$ the size 
of the GAIA sample of potential systems which might be discovered and measured, 
$2)$ GAIA's ability in revealing the existence of a possibly large number of 
systems which might be bearing rocky, perhaps habitable planets, and $3)$ the 
impact of GAIA's coplanarity measurements in multiple-planet systems on the 
theoretical models of formation and evolution of planetary systems.

\subsection{The size of the GAIA sample }


In our earlier work~\cite{lattanzi00}, we considered in the 
simulations single giant planets, in the mass range 
$0.1\leq M_\mathrm{J}\leq 5$, orbiting 1-$M_\odot$ stars with periods up to 
twice the mission duration, and placing the systems at increasing 
distances from our Sun. We parameterized our results in terms of the 
astrometric signal-to-noise ratio $\alpha/\sigma_\psi$ between the 
astrometric signature $\alpha$ and the single measurement error, which we set 
to $\sigma_\psi = 10$ $\mu$as (implying a final accuracy of 4 $\mu$as), 
value which applies to stars brighter than $V = 13$. 
In practice, this choice for the mass of the parent star encompasses
the spectral class range from $\sim$ F0 to early K type dwarf stars,
whose masses are within a factor of $\sim$ 1.5 that of the Sun.
This, in combination with the $V<13$ magnitude limit, translates 
into a distance cutoff for detection and accurate orbit estimation 
of $\sim$ 200 pc. To this distance, F-G-K type dwarfs dominate the star counts
at bright magnitudes, and within this horizon modern galaxy models~\cite{bienayme87}
predict some $3-5\times 10^5$ solar-type dwarfs available for investigation.
Knowing the stellar content in the solar neighborhood and the planetary 
frequency distribution, we can extrapolate a number of potential planetary 
systems within GAIA's detection horizon. Early estimates~\cite{marcy99}
yield an integral planetary frequency of $\sim 3-4\%$ 
for giant planets in the mass range 0.5-5 $M_\mathrm{J}$ 
orbiting within 3 AU from the parent star. By assuming a uniform planetary 
frequency distribution with orbital semi-major axis~\cite{lattanzi99}, 
then we can derive a {\it lower limit} to the number of giant planets GAIA 
would detect and measure at a given distance $d$ (in pc). In Table~\ref{nplan} 
we summarize the results. The values of N$_d$ at different distances 
correspond to actual {\it new detections}, 
once the fractions of detected giant planets in common 
with the overlapping semi-major axis regions at lower distances have 
been properly subtracted. 

\begin{table}[t]
\begin{center}
   \caption{Number of giant planets that could be revealed by GAIA (N$_d$), 
and fraction of detected planets having accurate orbital elements determined 
(N$_m$), as a function of increasing distance from the Sun ($\Delta d$). 
A uniform frequency distribution of 1.3\% planets per 1-AU bin is assumed}
\vskip 0.3cm
   \renewcommand{\arraystretch}{1.4}
   \setlength\tabcolsep{8pt}
   \begin{tabular}{ccccc}
       \hline\noalign{\smallskip}
       $\Delta d$ (pc) & $N_\star$  & $\Delta a$ (AU)
&  $N_{\rm d}$  & $N_{\rm m}$\\
       \noalign{\smallskip}
       \hline
       \noalign{\smallskip}
0-100 & $\sim$61\,000 & 1.3 - 5.3 & $> 1600$ & $> 640$\\ 
100-150 & $\sim$114\,000 & 1.8 - 3.9 & $> 1600$ & $> 750$\\
150-200 & $\sim$295\,000 & 2.5 - 3.3 & $> 1500$ & $> 750$ \\

      \hline\noalign{\smallskip}
   \end{tabular}
\label{nplan}
\end{center}
\end{table}

Therefore, the total number of giant planets 
GAIA could discover orbiting around normal stars within the distance limit of 
200 pc from the Sun is then {\it greater} than 4700, and roughly $50\%$ of 
the detected planets would have orbital parameters and masses good to $20-30\%$, 
or better. The statistical value of such a sample (comparable in size to 
that of the observing lists of the largest ground based surveys) 
would be instrumental for critical testing of theories on planet formation 
and evolution. 

\subsection{GAIA and the Habitable Zones}\label{zone}

\begin{figure}[th]
\centering
\includegraphics[width=.80\textwidth]{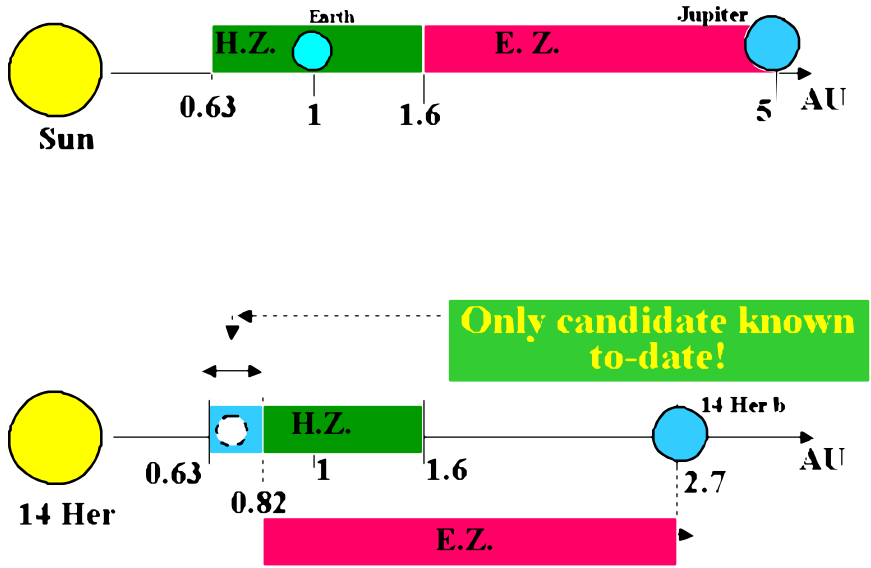}
\caption{Upper half: comparison between the Solar System's Habitable Zone 
(in green) and Exclusion Zone (in red). Jupiter's orbital period is sufficiently 
large that the entire Habitable Zone of the Sun is available for formation 
of rocky planets. Lower half: the same comparison in the case of the 14 Her 
system, the only one known to-date to bear a giant planet on a sufficiently large orbit 
for at least part of the Habitable Zone (in blue) to be able to host in 
principle a rocky planet} 
\label{habitable}
\end{figure}

With the current payload design~\cite{perryman01}, 
the range of planetary masses between 1 Earth-mass and a few 
Earth-masses (Neptune-class planets) will only be marginally accessible 
to GAIA's all-sky survey. Its astrometric accuracy will be sufficient 
to address the issue of their existence only around a handful of the closest 
stars, within 5-10 pc from the Sun. 

Nevertheless, GAIA's contribute to the 
search for rocky, possibly habitable planets will be significant. Theory, 
in fact, provides us with two important concepts: Habitable Zone and Exclusion 
Zone. The Habitable Zone~\cite{kasting93} is defined by the distance from a 
given star at which the temperature is such that water can be present in 
the liquid phase. The center of the Habitable Zone (whose distance depends 
on the mass of the central star) can be roughly identified by the formula 
$P/P_\oplus = (M/M_\odot)^{1.75}$. 
The Exclusion Zone~\cite{wetherill96} is defined by the 
dynamical constraint $P_\mathrm{G} > 6\times P_\mathrm{R}$, 
which states that for a rocky planet 
to form in the Habitable Zone of a star then a giant planet must form on an
orbital period $P_\mathrm{G}$ at least six times larger than the period 
$P_\mathrm{R}$ of the rocky planet. 
In Figure~\ref{habitable} we show these two concepts as they have been realized 
in our Solar System and in the only other interesting candidate known to-date, 
the 14 Her planet-star system.
All those systems GAIA will discover harboring a giant planet on a 
sufficiently wide orbit ($a\ge 3$ AU) would immediately be added to the list 
of targets for the next generation of missions which will search the Habitable 
Zones of such systems for evidence of the existence of terrestrial planets.

\subsection{GAIA and the planetary systems}

The observational evidence of the first extra-solar planetary systems, whose 
unexpected orbital configurations are very unlike the Solar System's, 
has immediately raised crucial question regarding their formation and 
evolution. Are the orbits coplanar? Are the configurations dynamically 
stable? Radial velocity measurements cannot determine either the 
inclination $i$ of the orbital plane with respect to the plane of the 
sky, or the position angle $\Omega$ of the line of nodes in the plane 
of the sky. General conclusions on the architecture, orbital evolution 
and long-term stability of the newly discovered planetary systems cannot 
be properly assessed without knowledge of the full set of orbital parameters 
and true mass values. 

GAIA will be capable of detecting and measuring a variety of configurations 
of potential planetary systems. Utilizing as a template the two outer 
planets in the $\upsilon$ Andromed\ae\, system, we have found~\cite{sozzetti01} 
how a 60--pc limit on distance holds for
detection and measurement accurate to 30\%, or better, of {\it
planetary systems} composed of planets with well-sampled periods
($P\leq 5$ yr), and with the smaller component producing 
$\alpha/\sigma_\psi\geq 2$. Also, accurate coplanarity tests, with relative 
inclinations measured to a few degrees, will be 
possible for systems producing $\alpha/\sigma_\psi\geq 10$~\cite{sozzetti01}.

The frequency of multiple-planet systems, and their preferred
orbital spacing and geometry are not currently known. Based on
star counts in the vicinity of the Sun extrapolated from modern
models of stellar population synthesis, constrained to bright
magnitudes ($V< 13$ mag) and solar spectral types (earlier than
K5), we should expect $\sim 13\,000$ stars to 60 pc~\cite{lattanzi99}. 
GAIA, in its high--precision astrometric survey of the
solar neighborhood, will observe each of them, searching for
planetary systems composed of massive planets in a wide range of
possible orbits, making accurate measurements of their orbital
elements and masses, and establishing quasi-coplanarity (or
non-coplanarity) for detected systems with favorable
configurations.

\section{How can GAIA achieve this?}

Two major issues can be singled out at the moment of defining the 
crucial steps which must be undertaken in order for GAIA 
to fully accomplish the scientific goals in the field of extra-solar 
planetary systems which have been outlined in the previous sections. 
First, specific requirements on the instrument 
performance must be met, and secondly it will be essential to identify 
the most robust and reliable procedures of analysis of the 
actual observational data. 

\subsection{The GAIA instrument}

\begin{figure}[tbh]
\centering
\includegraphics[width=.60\textwidth]{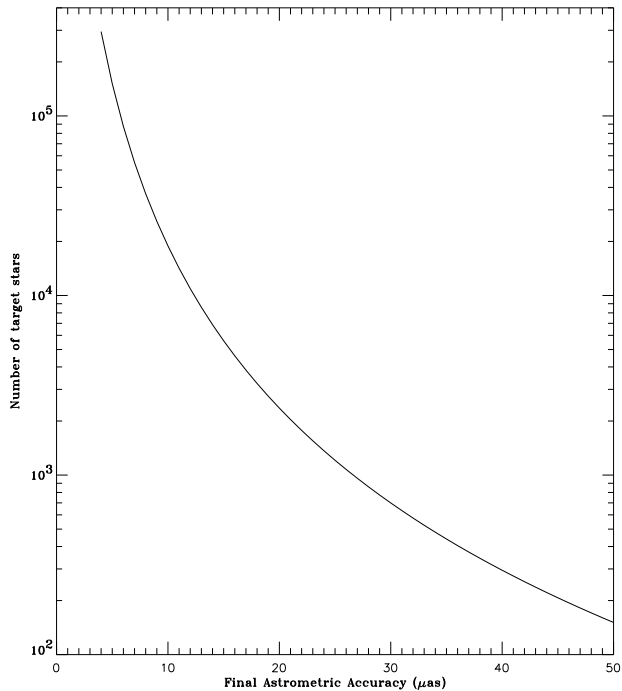}
\caption{Number of solar-type stars available for planet searches 
with GAIA, as function of the final accuracy on positions, proper 
motions, and parallaxes. If the final astrometric error is 8 $\mu$as 
instead of 4, then the size of the sample decreases by an order of 
magnitude ($\sim 2\times 10^4$ vs. $\sim 3\times 10^5$). In the limiting 
case of final astrometric accuracy equal to that of DIVA and FAME on 
bright objects, then the number reduces to some 150 stars)} 
\label{horizon}
\end{figure}

In order to fulfill the expectations for ground-breaking results in field 
of planetary science, the GAIA instrument~\cite{perryman01} 
must meet the stringent 
requirement of 4 $\mu$as final astrometric accuracy on positions, proper 
motions, and parallaxes for bright targets ($V < 13$). 
In fact, in order to keep the ratio $\alpha/\sigma_\psi = const.$, 
an increase in the measurement error implies 
an increase in the astrometric signature due to the planet. 
In particular, the same 
type of system (same stellar mass, same planet mass, same orbital period), 
as $\sigma_\psi$ increases, would be detectable at increasingly smaller distances. 
In Figure~\ref{horizon} we show how the number of stars available 
to GAIA for astrometric planet searches would decrease as a function 
of the increasing final error accuracy, assuming that the number of 
objects scale with the cube of the radius (in pc) of a sphere 
centered around the Sun. If $\sigma_\psi$ is increased by a factor 
2 the number of stars available for investigation would already be reduced 
by an order of magnitude. In the limit for final astrometric accuracy 
equal to the one foreseen for DIVA and FAME on bright objects (50 $\mu$as), 
only some 150 stars would probably fall within GAIA's horizon. The size 
of the sample would be so reduced that extra-solar planets would completely 
disappear from the GAIA science case. 

\subsection{Modeling the Observations}
 
In order to quantify the scientific impact of GAIA global astrometry 
measurements in the field of extra-solar planets extensive 
simulations have been used during the last few 
years~\cite{lattanzi99,lattanzi00,sozzetti01}.

Future work will concentrate on: $a)$ refinements of the models of 
observations and observables. In particular, the science object 
position as defined in the satellite's reference frame will be 
expressed as function of all possible kinematic and dynamical parameters and of 
all astrophysical effects which contribute to a significant motion of the stellar 
photocenter, at the level of the single-measurement error. To this aim, 
more realistic galaxy models will be used, together with detailed models 
of specific environments, which will be needed for example in the case of 
the search for planets in stellar associations; $b)$ as new knowledge is 
obtained from continuous improvement in the understanding of the instrument 
errors and performance before launch, a more realistic error model 
for GAIA observations will be implemented, which includes all possible 
sources of instrumental and 
astrophysical systematic errors, and their correlations; $c)$ for a proper 
assessment of the effectiveness of the overall search and optimization 
strategy, the analysis tools will have to be refined in order to obtain a realistic 
estimation process, which involves the implementation of refined models for 
global search and optimization strategies of starting guesses for the orbital 
parameters. To this end, actual ground-based (or simulated) spectroscopic 
data could be used jointly with the simulated astrometric dataset, to improve 
orbital solutions and determine the full three-dimensional motion of the 
analyzed systems.


\end{document}